\documentclass[twocolumn,showpacs,aps,prl,floatfix,%
superscriptaddress]{revtex4}  

\usepackage{graphicx} 

\newcommand{\ee}{\end{equation}} 
\newcommand{\eq}{{\,=\,}} 
\newcommand{\bm}{\bf}

\begin{document} 
 
 
\title{Di-jet hadron pair correlation in a hydrodynamical model with a quenching jet}
  
\date{\today}
  
\author{A. K. Chaudhuri} 
\email{akc@veccal.ernet.in} 
\affiliation{Variable Energy Cyclotron Centre, 1-AF, Bidhan Nagar,
Kolkata - 700 064, India} 
 
\begin{abstract}

In jet quenching, a hard QCD parton, before fragmenting into a jet of hadrons, deposits a fraction of its energy in the medium, leading to  suppressed production of high-$p_T$ hadrons. 
Assuming that the deposited  energy quickly thermalizes, we simulate the subsequent hydrodynamic 
evolution of the QGP fluid. Hydrodynamic evolution and subsequent particle emission depend on the jet trajectories. Azimuthal distribution of excess $\pi^-$ due to quenching jet, averaged over all the trajectories, reasonably well reproduce the di-hadron correlation as measured by the STAR and PHENIX collaboration in central and in peripheral Au+Au collisions.  
 \end{abstract} 
 
 
\maketitle 
 

  From the general theoretical considerations, it was
predicted \cite{QGP3jetqu} that in a dense deconfined medium, high-speed partons will suffer energy loss, significantly modifying the fragmentation function, which in turn will lead to suppressed production of hadrons. The phenomena called "jet quenching", was later verified at Relativistic Heavy Ion Collider (RHIC), in Au+Au collisions at $\sqrt{s}_{NN}$=200 GeV \cite{Adcox:2001jp,Adler:2002xw,Adams:2003kv}.
However it is unclear how the lost energy is transported in the dense medium. 
It has been suggested that a fraction of lost energy will go to collective excitation, call the
"conical flow" \cite{Stoecker:2004qu,shuryak,Casalderrey-Solana:2005rf}. The parton moves with speed of light, much greater
than the speed of sound of the medium ($c_{jet} >> c_s$), and the quenching jet can produce a shock wave with Mach cone angle, $\theta_M=cos^{-1}c_s/c_{jet}$. Resulting conical flow will have characteristic peaks at $\phi=\pi-\theta_M$ and $\phi=\pi + \theta_M$. Both in STAR \cite{Wang:2004kf,Adams:2005ph} and PHENIX \cite{Jacak:2005af} experiments,  indication of such peaks are seen in azimuthal distribution of secondaries associated with high $p_T$ trigger in central Au+Au collisions.
Mach like structure (splitting of away side peak) can also
be obtained in various other models, e.g. gluon Cerenkov like radiation models \cite{Dremin:2005an,Koch:2005sx}, the parton cascade model \cite{Ma:2006mz}, the Markovian parton scattering model \cite{Chiu:2006pu}, the color wake model
 \cite{Ruppert:2005uz}.
Recently in  \cite{Chesler:2007an} energy density wake produced by a heavy quark moving through 
a strongly coupled N=4 supersymmetric, Yang-Mills plasma is computed using ADS/CFT correspondence. Mach cone like structures is also observed for quark velocity greater than the speed of sound of the medium. 

Recently, we have numerically solved hydrodynamical equations with an (time dependent) source, representing the quenching jet \cite{Chaudhuri:2005vc,Chaudhuri:2006qk,Chaudhuri:2007gq}. It is assumed that the lost energy is quickly thermalised. Nevertheless, the energy is deposited locally along the trajectory of the fast parton, leading to local energy density  inhomogeneities, which if thermalised should in turn evolve hydrodynamically.  
Explicit simulation of hydrodynamic evolution with a quenching jet, indicate that the evolution of the fluid as well as subsequent particle emission are strongly influenced by the jet path length in the medium \cite{Chaudhuri:2005vc,Chaudhuri:2006qk,Chaudhuri:2007gq}. Evolution produces 'distorted'  Mach shock front like surfaces. As predicted in \cite{Satarov:2005mv}, finite fluid velocity
and also inhomogeneity of the fluid, distorts the Mach surfaces. Azimuthal distribution of pions, due to quenching jet only, do not show the characteristic of 'conical' flow. Rather, depending on the jet trajectory, it show a single peak either at $\phi >\pi$ or $\phi <\pi$. The distribution averaged over all the jet trajectories,
show two peaks with a dip at $\phi=\pi$,   mimicking the conical flow.    However, in 
\cite{Chaudhuri:2006qk}, we did not account for the limited $p_T$ range $(0.15 \leq p_T \leq 4)$ in the STAR experiment. Moreover,
in \cite{Chaudhuri:2006qk}, jet trajectories were  
characterized by a single parameter, $\phi_{prod}$, while in a two-dimensional calculation, as in \cite{Chaudhuri:2006qk}, unique characterization of jet trajectories need at least two parameters.  
In the present paper, we have corrected the deficiencies in the model.  Also,   in addition to the STAR data \cite{Wang:2004kf}, we analyze the  PHENIX data \cite{Jacak:2005af} on the di-hadron angular correlation in 0-5\%, 5-10\% and 60-90\% centrality Au+Au collisions. 

\begin{figure}[h] 
\includegraphics[bb=86 354 530 793 ,width=0.7\linewidth,clip]{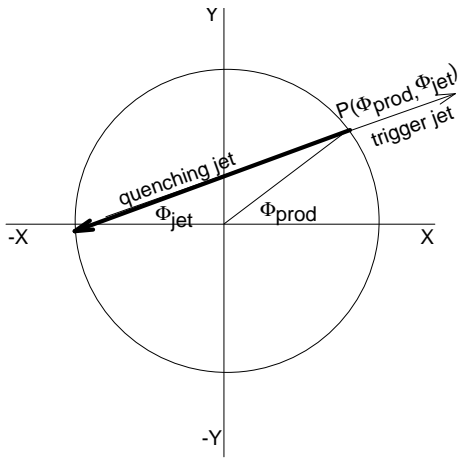}
\caption{Schematic representation of a jet moving through the
medium. The high $p_T$ pair is assumed to produce at $P$ on the surface of the fireball characterized by the angle
$\phi_{prod}$. One of the jet escapes forming the trigger jet, the
other move in the fireball at an angle $\phi_{jet}$.}
\label{F1}
\end{figure}

A schematic representation
of the jet moving through the medium is shown in Fig.\ref{F1}.
We assume that just before hydrodynamics become applicable, a di-jet 
of high-$p_T$ partons is produced.  
Strong jet quenching and survival of the trigger jet, forbids production
in the interior of the fireball. Jet pairs can be produced only on a thin
shell on the surface of the fireball. For Au+Au collisions at
impact parameter $b$, we assume that the di-jet is produced on  
the surface of the ellipsoid with minor and major axis, $A=R-b/2$ and
$B=R\sqrt{1-b^2/4R^2}$, with $R=6.4$fm.
One of the jet moves outward  and escapes, forming the trigger jet. The
other enters into the fireball. As shown in Fig.\ref{F1}, the trajectory
of the quenching jet can be uniquely characterized by
two angles, $\phi_{prod}$,
($-\pi \leq \phi_{prod} \leq \pi$) and 
$\phi_{jet}$, ($-\pi/2 \leq \phi_{jet} \leq \pi/2$).
The fireball is expanding and cooling. The ingoing parton travels 
at the speed of light and loses energy in the fireball which 
thermalizes and acts as a source of energy and momentum for 
the fireball medium.  
We solve the energy-momentum conservation equation,

\begin{equation} \label{1}
\partial_\mu T^{\mu\nu}=J^\nu,
\end{equation}

\noindent where the source is modeled as,

\begin{eqnarray} 
\label{2} 
 &&J^\nu(x)=J(x)\,\bigl(1,-cos(\phi_{jet}),-sin(\phi_{jet}),0\bigr),\\
\label{3}
 &&J(x) = \frac{dE}{dx}(x)\, \left|\frac{dx_{\rm jet}}{dt}\right| 
          \delta^3(\bm{r}-\bm{r}_{\rm jet}(t)).
\end{eqnarray}

Massless partons have light-like 4-momentum, so the current $J^\nu$
describing the 4-momentum lost and deposited in the medium by the 
fast parton is taken to be light-like, too. $\bm{r}_{\rm jet}(t)$ is 
the trajectory of the jet moving with speed $|dx_{\rm jet}/dt|\eq{c}$.
$\frac{dE}{dx}(x)$ is the energy loss rate of the parton as it moves 
through the liquid. It depends on the fluid's local rest 
frame particle density. Taking guidance from the 
phenomenological analysis of parton energy loss observed in Au+Au 
collisions at RHIC \cite{Eloss} we take
\begin{equation}
\label{4}
  \frac{dE}{dx} = \frac{s(x)}{s_0} \left.\frac{dE}{dx}\right|_0
\end{equation}
where $s(x)$ is the local entropy density without the jet. 
The measured suppression of high-$p_T$ particle production in Au+Au 
collisions at RHIC was shown to be consistent with a parton energy 
loss of $\left.\frac{dE}{dx}\right|_0\eq14$\,GeV/fm at a reference 
entropy density of $s_0\eq140$\,fm$^{-3}$ \cite{Eloss}. We do not
considered the possibility that some jets might be stopped in the
medium. It is implicitly assumed that the ingoing parton has enough energy to pass through the medium. The energy loss is weighted by the entropy density, as the fluid evolve, entropy density decreases and at the late stage of the evolution
energy loss is minimum.
 
We solve Eqs.\ref{1} in  
$(\tau=\sqrt{t^2-z^2},x,y,\eta{=}\frac{1}{2}\ln\left[\frac{t{+}z}{t{-}z}\right])$  coordinates, assuming boost-invariance. The source term Eq.\ref{3} violate the assumption of boost-invariance. We 
therefore modify it by replacing the $\delta$-function 
in (\ref{3}) by

\begin{eqnarray}
\label{5}
  \delta^3(\bm{r}-\bm{r}_{\rm jet}(t)) &\longrightarrow&
  \frac{1}{\tau}\,\delta(x-x_{\rm jet}(\tau))\,\delta(y-y_{\rm jet}(\tau))
\nonumber\\
 &\longrightarrow&\frac{1}{\tau} \, 
\frac{e^{-(\bm{r}_\perp-\bm{r}_{\perp,{\rm jet}}(\tau))^2/(2\sigma^2)}}
      {2\pi\sigma^2}
\end{eqnarray}
with $\sigma{\,=\,}0.70$\,fm, $r_\perp=(x,y)$. 

Intuitively, this replaces the ``needle'' 
(jet) pushing through the medium at one point by a ``knife'' cutting the
medium along its entire length along the beam direction. Rather than
'conical', the replacement will produce a 'wedge' flow,
  over estimating the
effect of jet quenching.

The hydrodynamical equations are solved with the standard 
initialization described in \cite{QGP3v2},   corresponding to a 
peak initial energy density of $\varepsilon_0 \eq 30$ $GeV/fm^3$ at
$\tau_0 \eq 0.6$ $fm/c$. We use the equation of state EOS-Q 
 described 
in \cite{QGP3v2} incorporating a first order phase transition 
and hadronic chemical freeze-out at a critical temperature 
$T_c{\,=\,}164$\,MeV. The hadronic sector of EOS-Q is soft with a 
squared speed of sound $c_s^2 \approx 0.15$. 

\begin{figure}[h] 
\includegraphics[bb=31 93 535 791
 ,width=0.7\linewidth,clip]{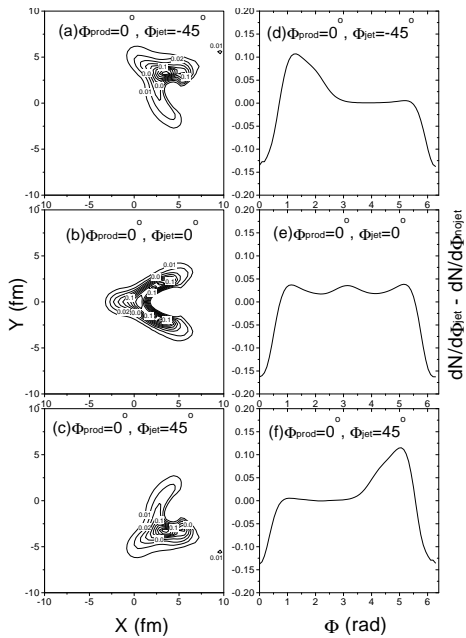}
\caption{ In left panels contour plot of excess energy density due to a quenching jet in b=2.3 fm Au+Au collisions are shown. The di-jet is produced at $\phi_{prod}=0^\circ$, but with different orientations; (a)   $\phi_{jet}=-45^\circ$, (b)$\phi_{jet}= 0^\circ$ and
(c)$\phi_{jet}=45^\circ$ respectively. In the right panels, azimuthal distribution 
of $\pi^-$, due the quenching jet only, from the corresponding evolution are shown.}
\label{F2}
\end{figure}

Some results of our simulations, for Au+Au collisions at
impact parameter b=2.3 fm,  are shown in Fig.\ref{F2}.   In left panels
 (a),(b) and (c), for a few jet trajectories, contour plots of 'excess'
 energy density (energy density in evolution with a quenching jet minus
 the energy density in evolution without a quenching jet), after 8 fm of
 evolution,  are shown.   The contour plots of 'excess' energy density clearly show that the
 hydrodynamic evolution of the QGP fluid, in presence of a quenching jet,
 depend on the
jet path length. For example, excess energy density distribution
in panel (a) and panel (c) are identical except for a  rotation about
$\pi/2$. For those two trajectories, the jets traverse a similar path 
length in the fluid.
We also note the excess energy density distribution
in all the jet trajectories  show Mach cone like surfaces.   One also
notices the distortion of Mach cone like surfaces. In panel (b)
the excess energy density distribution is symmetric with respect to
the jet axis,
but not in panel (a) or (c). Finite fluid velocity and inhomogeneity of
the medium distorts the Mach cone like surfaces 
\cite{Chaudhuri:2006qk,Satarov:2005mv}.

In the right panels (d), (e) and (f)  azimuthal distribution of
$\pi^-$ due to the quenching jet  are shown.
 Using the standard  
Cooper-Frey prescription, for each jet trajectory,  we have calculated
the azimuthal dependence    of $p_T$ integrated ($1 < p_T < 2.5 $GeV) $\pi^-$ yield $(\frac{dN}{d\phi})_{jet}$  at
freeze-out temperature, $T_F$= 100 MeV. Azimuthal angle $\phi$ is measured with respect to the quenching jet axis. We also calculate the $\pi^-$ yield $(\frac{dN}{d\phi})_{nojet}$ in an evolution with identical conditions but for the quenching jet.  
In (d)-(e) azimuthal distribution of    
excess pion yield $ (\frac{dN}{d\phi})_{jet}-(\frac{dN}{d\phi})_{nojet}$, are shown.
Excess $\pi^-$ distribution do not show a two peak structure, rather it show a single peak either at $\phi >\pi$ or at $\phi < \pi$. Only when the jet is along the diameter, one find a broad peak centered around $\phi=\pi$.
A similar picture is seen at other trajectories also. 
Indeed, we find that, depending on the trajectory, a quenching jet
produces peak either at  
$\phi_- \sim 2.\pm 1$ rad or at $\phi_+ \sim 4 \pm 1$ rad. As it will be shown later, when 
many events are summed up, a two peak structure, akin to 
conical flow, appear. The conical flow or the splitting of the away side jet is an average effect, not to be seen in
a single event. 

\begin{figure}[ht] 
\includegraphics[bb=36 391 440 793
 ,width=0.7\linewidth,clip]{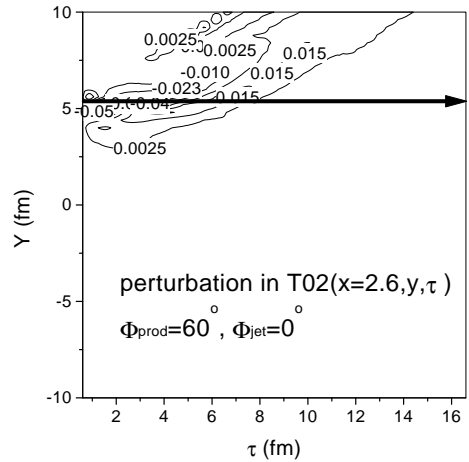}
\caption{contour plot of perturbation in momentum density $T^{02}$ ($T^{02}$ in evolution with a jet minus $T^{02}$ in evolution without a jet) in $\tau-y$ plane, at a fixed x=2.6 fm. The trajectory of the 
jet is shown by the black line with arrow. Perturbation in
$T^{02}$ do not remain confined along the jet trajectory.}
\label{F3}
\end{figure}

In Fig.\ref{F3}, we have shown the perturbation produced in the momentum density 
$\Delta T^{02} =T^{02}_{jet} (x,y)-T^{02}_{nojet} (x,y)$
due to a quenching jet. The trajectory of the jet is shown by the straight line. $\Delta T^{02}$  do not remain confined along the jet trajectory, it deviates sideward.
Even though the jet is restricted to move parallel to the x-axis and expected to produce a peak at $\phi=\pi$,  
due to finite fluid velocity and inhomogeneity, the momentum perturbation moves sideward and produce a peak at $\phi < \pi$. 
We may mention that $\Delta T^{02}$  rather look like
that due to a deflected jet \cite{Chiu:2006pu}.  
In the deflected jet picture  also , in a single jet event, the away jet is deflected sideward, resulting in a side ward peak. When many events are summed up, a double peak structure appears in the azimuthal distribution.  Two-particle angular correlation can not discriminate between the two scenarios, e.g. the distorted Mach shock wave or the deflected jet. 
Three-particle angular correlation can possibly discriminate between them.

\begin{figure}[ht] 
\includegraphics[bb=26 260 562 795
 ,width=0.75\linewidth,clip]{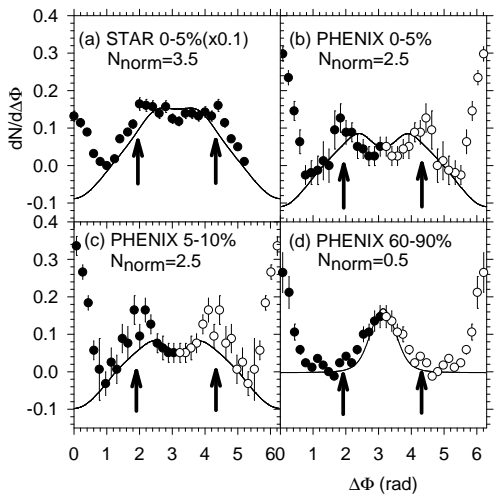}
\caption{Di-hadron correlation as measured by the  STAR \cite{Wang:2004kf}  and the PHENIX \cite{Jacak:2005af} collaboration.  
The filled circles in (b)-(d) are (original) PHENIX data, the blank circles are 
generated by reflection. The arrows in 4 panels indicate the approximate Mach peak position for average speed of sound 0.333 \cite{Casalderrey-Solana:2005rf}. The solid lines are the present (hydro+jet) model calculation for the azimuthal distribution of excess $\pi^-$  due to quenching jet, averaged over all possible quenching jets. They are normalized by a factor $N_{norm}$.}
\label{F4}
\end{figure}

In Fig.\ref{F4}, in 4 panels we have shown the di-hadron correlation as measured by the STAR  \cite{Wang:2004kf}  and the PHENIX \cite{Jacak:2005af} collaboration. STAR collaboration measured the correlation function in 0-5\% centrality Au+Au collisions. They measured charged hadrons ( $0.15 \leq p_T \leq 4 GeV$) associated with high $p_T$ ($4 \leq  p_T^{trigger} \leq 6 GeV$) trigger particle. The correlation function shown in Fig.\ref{F4}a, barely show the splitting of the away side peak.  
PHENIX collaboration measured the correlation function as a function of centrality of collisions. They constructed the correlation function between the trigger ($2.5 \leq p_T \leq 4 GeV$) and charged hadrons ($1 \leq p_T \leq 2.5 GeV$). PHENIX correlation function in 0-5\%, 5-10\% and 60-90\% centrality collisions are shown in Fig.\ref{F4}b,c and d.  
In 0-5\% and 5-10\% centrality collisions,
the away side peak is splitted into two peaks, but not in 60-90\% centrality collisions. In 60-90\% centrality collisions show
the usual structure in jet event, two peaks, one at $\phi=0$ and the other at $\phi=\pi$. We also note that , splitting of the away side jet is more prominent in PHENIX than in STAR experiment, presumably due to more hard hadrons in PHENIX measurements. 

In Fig.\ref{F4}, solid lines are the $p_T$ integrated azimuthal distribution of excess pions in Au+Au collisions at impact parameter b=2.3 fm (panel (a) and (b)), b=4.1 fm (panel(c)) and b=12.1 fm (panel (d)) Au+Au collisions. They roughly corresponds to 0-5\%, 5-10\% and
60-80\% centrality Au+Au collisions. The $\pi^-$ production was averaged over all possible jet trajectories. For each jet trajectory
($\phi_{prod}$, $\phi_{jet}$), we calculate the $p_T$ integrated 
excess pions $\frac{dN}{d\Delta\phi}(\phi_{prod},\phi_{jet})$ and average over all the possible jet trajectories,

\begin{widetext}
\begin{equation}
<\frac{dN}{d\Delta\phi}>=
\frac{1}{2\pi} \int^{\pi}_{-\pi} d\phi_{prod}  
\left[\frac{1}{\pi} \int^{\pi/2}_{-\pi/2} d\phi_{jet}  
\frac{dN}{d\Delta\phi}(\phi_{prod},\phi_{jet})\right]
\end{equation}
\end{widetext}

Azimuthal distribution of excess $\pi^-$, normalized by a factor $N_{norm}\approx 2.5-3.5$, reasonably well reproduces the shape of the di-hadron correlation in STAR and PHENIX measurements   in central Au+Au collisions. It must be emphasized, the present model do not contain any parameter, other than $N_{norm}$. Considering that  STAR and PHENIX measured charged hadron, of which only $\sim$ 70\% are charged pions, $N_{norm}\approx$  2.5-3.5, seems
reasonable.  
The subtle difference between the STAR and PHENIX 0-5\% centrality data (splitting of the away side peak is prominent in PHENIX but not in STAR measurement) is also well reproduced in the model. The difference is due to $p_T$ range of associated particles. STAR data contain more soft hadrons than in PHENIX data.  Hydrodynamics is not very successful in peripheral collisions. Still the model reproduces PHENIX data in 60-90\% centrality collisions.

Apparently present results are at variance with the conclusion of
\cite{Chaudhuri:2005vc}. In \cite{Chaudhuri:2005vc},
it was concluded that the quenching jet possibly can explain the broadening of the away side peak but  not the splitting.
The conclusion was based on a single jet trajectory along the diameter. Distortion of Mach cone like surfaces in off-diameter trajectories produces the splitting.  
Mechanism of splitting of the away side peak in central and mid-central Au+Au collisions can now be understood in terms of jet quenching. An individual jet event does not produce 'conical' flow.
Depending upon the trajectory, an individual jet produces associated particles either at 
$\phi_+ \pm \delta\phi > \pi$ or $\phi_- \pm \delta\phi < \pi$.  
When averaged over all the jet trajectories, the associated particles
show a two peaks structure, much akin to the 'conical' flow. 
However, we also notice that though the model correctly predict the splitting of the away side jet in to two peaks, the peaks in experiments are  sharper than in the model calculations. The reason is possibly the assumption of boost-invariance. The 'needle' like jet is replaced by a 'knife', flattening the peaks.  
The model calculations also  do not reproduce the exact the experimental peak positions. Simulated peaks are   shifted 
from the experiment  by about $\sim 0.4$ rad. As shown in  
\cite{Chaudhuri:2006qk} the peak in the azimuthal distribution of associated particles, depend,  non-trivially, on the equation of state (speed of sound of the) medium. Deviation reflect our far from satisfactory    knowledge of  equation of state of the quarks matter for  of the QGP/hadronic matter.


\end{document}